\documentclass[prc,preprint,showpacs,tightenlines,%
                        amsfonts,nofootinbib]{revtex4}

\usepackage{graphicx}  %
\usepackage{bm}  %
\usepackage{amssymb}

\begin{document}

%
%

\newcommand{\alphabar}{\overline\alpha}
\newcommand{\alphabarp}{\overline\alpha\,{}'}
\newcommand{\Azero}{A_0}

\newcommand{\bzero}{b_0}

\newcommand{\Cs}{C_{\rm s}}
\newcommand{\cs}{c_{\rm s}}
\newcommand{\Cv}{C_{\rm v}}
\newcommand{\cv}{c_{\rm v}}

\newcommand{\dalem}{\frame{\phantom{\rule{8pt}{8pt}}}}
\newcommand{\del}{\partial}
\newcommand{\delN}{{\wt\partial}}
\newcommand{\Deltaevac}{\Delta{\cal E}_{\rm vac}}

\newcommand{\ed}{{\cal E}}
\newcommand{\edk}{{\cal E}_k}
\newcommand{\edkzero}{{\cal E}_{k0}}
\newcommand{\edv}{{\cal E}_{\rm v}}
\newcommand{\edvphi}{{\cal E}_{{\rm v}\Phi}}
\newcommand{\edvphizero}{{\cal E}_{{\rm v}\Phi 0}}
\newcommand{\edzero}{{\cal E}_{0}}
\newcommand{\Efermi}{E_{{\scriptscriptstyle \rm F}}}
\newcommand{\Efermistar}{E_{{\scriptscriptstyle \rm F}}^\ast}
\newcommand{\Efermistarzero}{E_{{\scriptscriptstyle \rm F}0}^\ast}
\newcommand{\etabar}{\overline\eta}
\newcommand{\ezero}{e_0}

\newcommand{\fomega}{f_\omegav}
\newcommand{\fpi}{f_\pi}
\newcommand{\fv}{f_{\rm v}}
\newcommand{\fvt}{\widetilde\fv}

\newcommand{\gA}{g_A}
\newcommand{\gammazero}{\gamma_0}
\newcommand{\GH}{\GKS}
\newcommand{\GKS}{G_{\rm KS}}
\newcommand{\gomega}{g_\omegav}
\newcommand{\gpi}{g_\pi}
\newcommand{\grho}{g_\rho}
\newcommand{\grad}{{\bm{\nabla}}}
\newcommand{\gs}{g_{\rm s}}
\newcommand{\gv}{g_{\rm v}}

\newcommand{\fm}{\mbox{\,fm}}

\newcommand{\infm}{\mbox{\,fm$^{-1}$}}
\newcommand{\isovectorTensor}{s_{\tauvec}}
\newcommand{\isovectorTensorN}{\wt\isovectorTensor}
\newcommand{\isovectorVector}{j_{\tauvec}}
\newcommand{\isovectorVectorN}{\wt j_{\tauvec}}

\newcommand{\kappabar}{\overline\kappa}
\newcommand{\Kbar}{\skew3\overline K}
\newcommand{\kfermi}{k_{{\scriptscriptstyle \rm F}}}
\newcommand{\kfermizero}{k_{{\scriptscriptstyle \rm F}0}}
\newcommand{\Kzero}{K_0}

\newcommand{\lagrang}{{\cal L}}
\newcommand{\LdotS}{\bm{\sigma\cdot L}}
\newcommand{\lsim}{\lower0.6ex\vbox{\hbox{$\ \buildrel{\textstyle <}
         \over{\sim}\ $}}}
\newcommand{\lzero}{l_{0}}

\newcommand{\Mbar}{\overline M}
\newcommand{\Mbarzero}{\Mbar_0}
\newcommand{\MeV}{\mbox{\,MeV}}
\newcommand{\momega}{m_\omegav}
\newcommand{\mpi}{m_\pi}
\newcommand{\mrho}{m_\rho}
\newcommand{\ms}{m_{\rm s}}
\newcommand{\Mstar}{M^\ast}
\newcommand{\Mstarzero}{M^\ast_0}
\newcommand{\mv}{m_{\rm v}}
\newcommand{\mzero}{{\rm v}_{0}}

\newcommand{\Nbar}{\skew3\overline N}

\newcommand{\omegaV}{V}
\newcommand{\omegav}{{\rm v}}

\newcommand{\Phizero}{\Phi_0}
\newcommand{\psibar}{\overline\psi}
\newcommand{\psidagger}{\psi^\dagger}
\newcommand{\pvec}{{\bf p}}

\newcommand{\rhoB}{\rho_{{\scriptscriptstyle \rm B}}}
\newcommand{\rhoBt}{\wt\rho_{{\scriptscriptstyle \rm B}}}
\newcommand{\rhoBzero}{\rho_{{\scriptscriptstyle \rm B}0}}
\newcommand{\rhominus}{\rho_{-}}
\newcommand{\rhoplus}{\rho_{+}}
\newcommand{\rhos}{\rho_{{\scriptstyle \rm s}}}
\newcommand{\rhospzero}{\rho'_{{\scriptstyle {\rm s} 0}}}
\newcommand{\rhost}{\wt\rho_{{\scriptstyle \rm s}}}
\newcommand{\rhoszero}{\rho_{{\scriptstyle {\rm s}0}}}
\newcommand{\rhotau}{\rho_{\tauvec}}
\newcommand{\rhotaut}{\wt\rho_{\tauvec}}
\newcommand{\rhothree}{\rho_{3}}
\newcommand{\rhothreet}{\wt\rho_{3}}
\newcommand{\rhozero}{\rho_0}

\newcommand{\scalar}{\rhos}
\newcommand{\scalarN}{{\rhost}}
\newcommand{\Szero}{S_0}

\newcommand{\tauvec}{{\bm{\tau}}}
\newcommand{\tauthree}{\tau_3}
\newcommand{\tr}{{\rm tr\,}}
\newcommand{\Tr}{{\rm Tr\,}}

\newcommand{\umu}{u^\mu}
\newcommand{\Ualpha}{U_{\alpha}}
\newcommand{\Ubar}{\skew2\overline U}
\newcommand{\Ueff}{U_{\rm eff}}
\newcommand{\Uzero}{U_0}
\newcommand{\Uzerop}{U_0'}
\newcommand{\Uzeropp}{U_0''}

\newcommand{\vecalpha}{{\bm{\alpha}}}
\newcommand{\veccdot}{{\bm{\cdot}}}
\newcommand{\vecnabla}{{\bm{\nabla}}}
\newcommand{\vecpi}{{\bm{\pi}}}
\newcommand{\vectau}{{\bm{\tau}}}
\newcommand{\vectorj}{j_{\scriptscriptstyle V}}
\newcommand{\vectorN}{{\wt\vector}}
\newcommand{\vecx}{{\bf x}}
\newcommand{\Vopt}{V_{\rm opt}}
\newcommand{\Vs}{V_s}
\newcommand{\Vzero}{V_0}

\newcommand{\wt}{\widetilde}
\newcommand{\wzero}{w_0}
\newcommand{\Wzero}{W_0}

\newcommand{\xvec}{{\bf x}}

\newcommand{\zetabar}{\overline\zeta}
%
%
%
%


\newcommand{\beq}{\begin{equation}}
\newcommand{\eeq}{\end{equation}}
\newcommand{\beqa}{\begin{eqnarray}}
\newcommand{\eeqa}{\end{eqnarray}}


\title{Covariant RPA in Effective Hadronic Field Theory}

\author{R. J. Furnstahl}\email{furnstahl.1@osu.edu}
\affiliation{Department of Physics,
         The Ohio State University, Columbus, OH\ \ 43210}
\author{J. Piekarewicz}\email{jorgep@csit.fsu.edu}
\affiliation{Department of Physics,
         Florida State University, Tallahassee, FL\ \ 32306}
\author{Brian D. Serot}\email{serot@iucf.indiana.edu}
\affiliation{Department of Physics and Nuclear Theory Center\\
             Indiana University\\ Bloomington, IN\ \ 47405\\}

%
\date{May, 2002}

\begin{abstract}
In an effective hadronic theory constructed to describe
long-range nuclear physics, the dynamics of the vacuum can 
be expanded in terms with zero or a finite number of
derivatives acting on the fields.
Thus vacuum dynamics can {\em always} be absorbed in the 
(infinite number of) counterterm
parameters necessarily present in the effective lagrangian.
These finite parameters, which at present must be fitted 
to data, encode the empirical vacuum physics
as well as other short-range
dynamics into the effective lagrangian; in practice,
only a small number of parameters must be fitted.
The strength of the effective field theory (EFT) framework
is that there is no need to make a
concrete picture of the vacuum dynamics, as one does in a 
renormalizable hadronic theory.
At the one-loop level, the most convenient renormalization
scheme requires explicit sums over long-range (``valence'') 
nucleon orbitals only, thus explaining the so-called ``no-sea
approximation'' used in successful covariant mean-field 
theory (MFT) calculations of static ground states.
When excited states are studied in the random-phase 
approximation (RPA), the same EFT scheme
dictates the inclusion of {\em both}
familiar particle-hole pairs {\em and} contributions that mix
valence and negative-energy single-particle Dirac wave functions.
The modern EFT strategy therefore justifies and explains
the omission of some explicit contributions from
the negative-energy Dirac sea of nucleons, as was 
done  to maintain conservation laws
in earlier pragmatic calculations of the nuclear
linear response.
\end{abstract}

\smallskip

\maketitle

\section{Introduction}
\label{sec:introduction}

Whereas RPA calculations of inelastic states
in finite nuclei using the simplest version of
covariant quantum hadrodynamics (QHD) have long been
available \cite{FURNSTAHL85,NISHIZAKI86,WEHRBERGER88,%
BLUNDEN88,SHEPARD89,DAWSON90,HOROWITZ90,PRICE92,WEHRBERGER93}, 
it is only in recent years that calculations based on 
accurately calibrated mean-field
theories have been 
performed~\cite{PIEKAREWICZ00,PIEKAREWICZ01,RING00,RING01}. 
This renewed
interest in covariant RPA, motivated in part by improved measurements
of various nuclear compressional
modes~\cite{DAVIS97,YOUNGBLOOD99,CLARK01}, leads us to examine these
RPA calculations within the context of effective field theory (EFT) 
descriptions of nuclear many-body systems,
which form the basis for modern QHD~\cite{SEROT97,FURNSTAHL00}. 

The basic issue involves the treatment of the quantum vacuum.
In theories where nucleons are described by four-component Dirac
spinors, one must consider the role of the Dirac sea of negative-energy
states.
In older calculations based on renormalizable QHD models, the 
mean-field (or one-baryon-loop) contributions to the
ground state from the Dirac sea could be calculated explicitly, 
but it was found that
these contributions precluded an accurate description of bulk nuclear
observables~\cite{SEROT86,SEROT92,FURNSTAHL97b,FURNSTAHL98}.
Therefore, for completely pragmatic reasons, these contributions were
omitted (in both renormalizable and nonrenormalizable models), resulting
in what has historically been called the ``no-sea approximation'' for
the mean-field ground 
state~\cite{BROCKMANN77,BROCKMANN78,RUFA88,REINHARD89,GAMBHIR90,RING96,RING01a}.
The ``no-sea'' hamiltonian, which typically contains sums over occupied
valence (positive-energy) Dirac wave functions and polynomials in
the mean meson fields, yields accurate results for bulk nuclear
observables~\cite{FURNSTAHL87a,REINHARD89,GAMBHIR90,%
RING96,FURNSTAHL97,SEROT97}.

Questions arose, however, in the treatment of the collective linear
response (RPA) of these ground states to external probes.
Whereas one might expect (based on the ``no-sea approximation'')
that this response would contain only the
well-known particle-hole contributions~\cite{MA97,MA97a,MA99}, 
it has long been known that
one must also include contributions from {\em negative-energy\/}
basis states, if fundamental principles such as Lorentz covariance and
gauge invariance are to be maintained~\cite{DAWSON90}.
This somewhat confusing situation is resolved in the modern EFT
approach~\cite{FURNSTAHL95,FURNSTAHL97}, 
which shows that the term ``no-sea'' is in fact a 
{\em misnomer}, and that consistent descriptions
of both the mean-field ground state and its linear response follow
naturally from the standard rules of quantum field theory.
The purpose of this paper is to illustrate these ideas as clearly as
possible.

Our most important conclusion is that in the EFT, 
{\em nothing} is omitted in the so-called ``no-sea approximation''
from either the ground state or the RPA linear response.
Although the old-fashioned interpretation discussed above implies that
(regulated) 
negative-energy loop contributions are {\em neglected} in the former
and that only the ``Pauli blocking'' corrections to the vacuum response
are included in the latter, EFT shows that this 
interpretation is {\em incorrect}.
In fact, the negative-energy contributions are always included, 
as one would expect from the rules of field theory, 
but they must be combined with the complete set of counterterms 
present in the QHD lagrangian; 
\emph{only the sum contributes to physical observables.}
Thus, even though the simple picture of the vacuum as a
negative-energy Dirac sea is likely to be incorrect
(given the complex nature of the QCD vacuum),
it is automatically corrected by combining the baryon loops (which are
well defined with a cutoff, for example) with the counterterms
and by fitting the resulting (unknown)
constants to empirical bulk nuclear 
properties~\cite{FURNSTAHL95,FURNSTAHL97b}.
In principle, these constants could be calculated directly from QCD.

This is the strength of the EFT: by fitting a small number of empirical
constants, we encode the correct vacuum dynamics into the mean-field
hamiltonian, {\em and there is no need to rely on a specific model
for the vacuum dynamics}, which is beyond the realm of the low-energy
EFT anyway.
Different renormalization/subtraction schemes shift contributions
between 
baryon loops and
counterterms without changing physical observables.
A particularly convenient prescription for nuclear ground states
implicitly cancels the sum over negative-energy states.
This procedure is equivalent to the so-called 
``no-sea approximation'', as we illustrate below.
Precisely the same renormalization scheme
(i.e., the counterterm parameters remain unchanged) 
must be applied to the linear
response, which leads automatically
to all required terms and maintains all important 
conservation laws.

\section{Covariant Effective Field Theory}
\label{sec:EFT}

An effective field theory (EFT) describes low-energy physics with
low-energy degrees of freedom. 
In some theories, like the Standard Model of Electroweak interactions,
the short-range (high-energy) contributions can be explicitly re-expressed
as terms in the low-energy, effective lagrangian.
In other effective field theories, like chiral perturbation theory or QHD,
the low-energy lagrangian cannot (yet) be calculated explicitly from QCD,
and the parameters must be fitted to experimental 
data~\cite{GEORGI93,FURNSTAHL95,FURNSTAHL97,VANKOLCK99}.

Guidance in choosing the form of a hadronic EFT comes from requiring that the
lagrangian maintain the symmetries of the underlying theory of QCD.
One also wants to choose an efficient set of low-energy degrees of freedom
(``generalized coordinates'') in the EFT lagrangian, to simplify the
treatment of the desired many-body problems.
Fortunately, in most applications of chiral perturbation theory or
descriptions of bulk nuclear structure, only a small number of parameters
are needed, and predictive power is 
retained~\cite{RING96,SEROT97,VANKOLCK99,FURNSTAHL00a,PIEKAREWICZ00,PIEKAREWICZ01}.
Thus no attempt is made in these EFTs to construct a detailed dynamical
description of the short-distance or vacuum physics.

The important point is that while the short-distance, ultraviolet behavior
of the effective theory is (probably) incorrect, it can be corrected
{\em systematically} by the normalization (or renormalization) of local
operators (``counterterms''), which have at most a finite number of
derivatives acting on the fields.
We emphasize that this procedure is {\em not} a prescription or a model for
describing the vacuum dynamics; we are truly encoding the appropriate
physics by fitting the unknown constants to data, using a lagrangian that
contains all (nonredundant) terms allowed by the underlying 
symmetries~\cite{FURNSTAHL97,SEROT97,VANKOLCK99}.

The ``no-sea approximation'' for the static, mean-field nuclear 
ground state can be understood as a
particularly economical way to define and choose the counterterms, 
although other, less efficient prescriptions could be made.
We review the arguments underlying this procedure 
below~\cite{FURNSTAHL95}.
Moreover, because the {\em same\/} 
counterterm parameters determine the linear response of the ground state
to external, time-dependent perturbations, a framework that manifests
their role will automatically produce a correct treatment of the RPA.

It is convenient to use an effective-action formalism to carry out the
EFT program at finite density
and to trace the role of the counterterms. The fundamental
object is the {\em effective action} 
$\Gamma[\phi,V^\mu]$ with spacetime dependent, classical,
Lorentz scalar and four-vector fields $\phi(x)$ and $V^\mu(x)$. 
$\Gamma[\phi,V^\mu]$ is obtained by a Legendre
transformation of the path-integral generating functional 
for propagators, which contains
sources coupled to the meson
fields~\cite{COLEMAN73,ILIOPOULOS75,ITZYKSON80,FURNSTAHL89,%
PERRY86,PERRY87,FURNSTAHL95,PESKIN95,WEINBERG96}.  
When evaluated with appropriate time-independent fields, $\Gamma$ is
proportional to the ground-state energy~\cite{ILIOPOULOS75,FURNSTAHL89}.  
It also generates the
one-particle-irreducible Green's functions and is therefore
related to the linear response of the system to external probes.  
Thus we can address the computation of
the ground state and the excited states in the same framework.  
For simplicity, we show only isoscalar, scalar and vector fields; 
the extension to other boson fields is straightforward but not
important for our discussion.
 
We consider only the one-loop effective action,
which generates the conventional mean-field or Hartree 
equations for the ground
state and the RPA equations for collective excited states.
This would seem to be a severe restriction.
Indeed, the successes of QHD mean-field theory are at first somewhat
mysterious, since the one-loop
approximation is just the finite-density
counterpart of the Born approximation at zero density, which
is inadequate for a quantitative description of nucleon--nucleon scattering.
However, density functional theory (DFT) can explain the successes
of these calculations and also provides a basis for understanding
the expansion and truncation of the QHD lagrangian.

Conventional DFT is based on an energy functional of the 
ground-state density of a many-body system, whose extremization
yields a variety of ground-state properties.
In a covariant generalization of DFT applied to nuclei, the
energy and grand potential
become functionals of the ground-state scalar density and
baryon-number four-current density.
Relativistic mean-field theories based on EFT are analogs of
the Kohn--Sham formalism of 
DFT~\cite{KOHN65,DREIZLER90,KOHN99,ARGAMAN00},
with local scalar and vector fields appearing in the role of
Kohn--Sham potentials~\cite{SEROT97}.
They are \emph{not} Hartree calculations using interactions designed
to reproduce free-space nucleon--nucleon observables.

Instead,
the one-loop energy [see Eq.~(\ref{eq:noseaE}) below] approximates the
exact energy functional, {\em which includes all higher-order
correlations}, using powers and gradients of 
auxiliary meson fields 
(or nucleon densities~\cite{FRIAR96,NIKOLAUS97,RUSNAK97}).
Multi-loop contributions are implicitly expanded in a generalized 
local-density approximation plus gradient corrections;
the success of this approach in Coulomb systems is well 
documented~\cite{DREIZLER90,KOHN99,ARGAMAN00}.
The level of truncation for the desired accuracy is
determined by EFT power 
counting~\cite{FURNSTAHL97,FURNSTAHL00a}.
This approximation is very accurate for the density regime
of interest, as verified by the excellent reproduction of
nuclear ground-state densities and
energies~\cite{GMUCA92,GMUCA92a,FURNSTAHL97,RUSNAK97}.
(For a more complete discussion, see sec.~6.1 of
Ref.~\cite{SEROT97} and Ref.~\cite{FURNSTAHL02}.)

The DFT also implies that we have a meaningful power counting
for the approximate calculation of the effective action, which
allows us to truncate the one-loop energy functional to any 
desired accuracy.  Thus our EFT is systematic to the extent that
the one-loop form of the energy functional is flexible enough
to be a good 
approximation to the most important multi-loop corrections.
Nevertheless, a fully systematic EFT expansion, in which loop
corrections can be included order-by-order in a small expansion
parameter, has yet to be developed.  We will return at the end
to discuss explicit improvements to the energy 
functional~\cite{HU00}.

To carry out the effective-action analysis of vacuum contributions, 
we start with the following 
lagrangian (density)
\beqa
  \lagrang (x) &=& \Nbar ( i \gamma^\mu \partial_\mu
       - \gv \gamma^\mu V_\mu - M +\gs \phi + \cdots ) N
       - \frac{1}{4}\, F_{\mu\nu} F^{\mu\nu}
       \nonumber \\
& & \quad 
      + \frac{1}{2}\, \mv^2 V^\mu V_\mu 
      + \frac{1}{2}\, \partial_\mu \phi \,\partial^\mu \phi
      - U ( \phi ) + \cdots \ ,
       \label{eq:lagrangian}
\eeqa
where $\gs$ ($\gv$) is the scalar (vector) coupling to
the nucleon, the vector field-strength tensor is 
$F_{\mu\nu} \equiv \partial_\mu V_\nu - \partial_\nu V_\mu$,
and $U (\phi )$ is an infinite polynomial in $\phi$.
The ellipsis represents contributions from other bosons (e.g.,
pions), a polynomial in (even powers of) the vector field,
a ``mixed'' polynomial involving both the scalar and vector
fields, and terms involving derivatives of the fields,
all of which are superfluous for the present illustration.
Moreover, the ellipsis contains the {\em counterterms},
which include all possible (nonredundant) terms allowed by
the symmetries of the theory; in particular, there are 
counterterm polynomials in the boson fields with exactly the
same form as those mentioned above.
There are also Lorentz-covariant counterterms involving the
nucleons (e.g., a wave function renormalization), 
which are needed when one calculates explicitly beyond one-loop
order~\cite{FURNSTAHL89,HU00}.

For a given approximation to the effective action, the counterterm
parameters can be fixed by any sufficiently
complete set of observables, and then
the \emph{same} parameters must apply to all calculations using the
effective action.
(This is equivalent to the emphasis in conventional RPA discussions
on using a consistent
interaction for the ground state and excited
states~\cite{DAWSON90,PIEKAREWICZ00,PIEKAREWICZ01}.)
Fitting to ground-state properties is predictive for
excited states that do not rely on unconstrained parameters or
on poorly approximated correlations.  Therefore, we expect that
{\em collective} excited states will be described well.

Since we expect the vacuum baryon-loop contributions to be largely 
canceled by the counterterms~\cite{FURNSTAHL97b}, 
it is efficient to make a reference subtraction to
build in this cancellation implicitly and to include (and fit) only 
the residual counterterms explicitly.
We can identify the subtraction 
by formally considering the effective action at zero
temperature and density (which is \emph{not} meant to describe
free-space scattering). 
The lagrangian enters in an exponential in a path integral
over all the fields, so we can start by integrating out the baryon
fields.
The boson fields act as auxiliary fields and can be redefined
(if necessary)
to eliminate any terms that are not bilinear in the baryon
fields;\footnote{%
This assumes that one has a useful power-counting expansion
of the effective lagrangian, so
that the undesired terms can be eliminated order-by-order
in the relevant parameter(s).}
thus, the result of the integration is a fermion 
determinant that contributes to the meson action as an additive
term given by~\cite{SEROT86,FURNSTAHL95}
\beq
  S_{\rm FD} [ \phi , V_\mu ] \equiv
     \int\! d^4x \, \lagrang_{\rm FD}
  = -i \,\Tr\! \ln K (0) \ ,
\eeq
where ``Tr'' indicates a trace over spacetime, spin, and
isospin.
The kernel $K (\mu )$ is defined by
\beq
    -i\, \Tr\!\ln K(\mu ) \equiv 
     -i\, \Tr\!\ln(i \gamma^\mu \partial_\mu
               + \mu\gamma^0
               - \Mstar - \gv \gamma^\mu V_\mu) 
               \ ,
\eeq
with the shorthand $\Mstar \equiv M - \gs \phi$, and the
chemical potential $\mu$ is introduced in this definition for
later convenience.
(Baryon counterterms that are needed beyond one-loop order
are suppressed.)
At present, we are working with $\mu = 0$, and we will comment
on this choice below.
Note that no approximation has been made at this point; 
$S_{\rm FD}$ is a \emph{functional} of the dynamical fields $\phi$ and
$V_\mu$ that must still be integrated over in the path 
integral.

The determinant can be evaluated using a derivative expansion
of the fields, which takes the form~\cite{PERRY86,PERRY87,FURNSTAHL90}
\beqa
  -i\, \Tr\!\ln K(0) &=& -i\, \Tr\!\ln(i \gamma^\mu \partial_\mu
               - \Mstar - \gv \gamma^\mu V_\mu ) \nonumber \\
    &=&
     \int\! d^4x \, [-U_{\rm eff}(\phi) 
        + \frac{1}{2}Z_{1s}(\phi)\partial_\mu\phi\,\partial^\mu\phi
         + \frac{1}{2}Z_{2s}(\phi)(\square\phi)^2 
         \nonumber \\
   & & \quad
         + \frac{1}{4}Z_{1v}(\phi)F_{\mu\nu} F^{\mu\nu}
         + \frac{1}{2}Z_{2v}(\phi)(\partial_\alpha F^{\alpha\mu})
              (\partial^\beta F_{\beta\mu})
         + \cdots \,] \ .
              \label{eq:Kzero}
\eeqa
This expansion reveals that the determinant contains all powers
of the boson fields and their derivatives that are allowed by
the symmetries.
For example, $U_{\rm eff}(\phi)$ is an infinite polynomial in 
$\phi$, and the conservation of baryon number implies that the
vector field can enter only in the combination $F_{\mu\nu}$.
This expansion was investigated in renormalizable QHD models and found
to be rapidly convergent~\cite{PERRY86,PERRY87,FURNSTAHL90}.

The contributions from $\Tr\!\ln K(0)$ represent vacuum
physics described by the Dirac sea of nucleons, which is
most likely incorrect.
The inadequacy of such a simple picture for the QCD vacuum
is demonstrated by the unnaturalness of
the coefficients in $U_{\rm eff}(\phi)$ under conventional QHD
renormalization \cite{FURNSTAHL97b}. 
Nevertheless,
the important point is that these vacuum contributions can be
written in terms of {\em local} products of fields and 
their derivatives that have the same form as the counterterms
already present in the meson lagrangian.
Thus the contributions from the fermion determinant can be
{\em exactly canceled} by the counterterms, leaving only the
original polynomial terms shown in Eq.~(\ref{eq:lagrangian}).
Since the remaining terms contain all possible forms allowed
by the symmetries, \emph{we can encode the true vacuum dynamics
into the lagrangian (or hamiltonian) by fitting the remaining
parameters to experimental data}.

To see how this works in practice, let us focus on the 
nonderivative term in Eq.~(\ref{eq:Kzero}), which is obtained by
treating the meson fields as constants.
(The gradient terms can be handled analogously.)
The nonderivative part of $\lagrang_{\rm FD}$ is an
infinite polynomial in $\phi$; at the one-loop level, one
finds:
\beqa
  \lagrang_{\rm FD} [ \phi ] &=& 
     i \int \frac{d^{\,\tau} k}{(2 \pi )^4} \ \tr \ln G^0 (k)
     \nonumber \\[5pt]
& & \quad {} + i \sum_{n\, = \, 1}^{\infty}
    \frac{(-1)^n}{n}\, [\gs \phi ]^n 
     \int \frac{d^{\,\tau} k}{(2 \pi )^4} \ \tr [G^0 (k)]^n \ .
\eeqa
Here the noninteracting baryon propagator is
$G^0 (k) = [\gamma^\mu k_\mu -M + i \epsilon ]^{-1}$,
``tr'' denotes a trace over spin and isospin,
and we have regularized dimensionally to maintain Lorentz
covariance and other symmetries.
One could also make these integrals finite by using an
explicit cutoff.
The point is that this polynomial (which is valid for {\em any}
values of the background field $\phi$) can be canceled exactly
by the implied counterterms in Eq.~(\ref{eq:lagrangian}),
leaving only the explicit polynomial $U (\phi )$.
Indeed, physical observables depend only on the {\em sum}
of all terms with the same structure, and so in practice,
{\em there is no need to compute explicitly either the loop
contributions or the counterterms}.
One simply considers the cancellations to be implicit and
takes the original potential $U (\phi )$ to contain finite,
renormalized parameters, which can ultimately be fitted to
empirical data to encode the true QCD vacuum dynamics into
the EFT.
Although in principle, an infinite number of meson terms
is needed to describe these effects, the principles of EFT
power counting and naturalness, which are validated by
phenomenological studies~\cite{FURNSTAHL97,FURNSTAHL00a},
show that one can truncate the lagrangian to a small
number of derivative and nonderivative terms in applications
to the structure of nuclei.
Thus only a small number of unknown constants must be
fitted to data to achieve accurate results, and the 
predictive power of the QHD effective action is maintained.

Now that the counterterm subtractions have been defined to
cancel the vacuum-loop contributions at zero density
and temperature, what happens at finite density or in a 
finite nucleus?
For these systems, we invoke the grand canonical ensemble and
allow $\mu$ to be nonzero.
(For simplicity, we continue to maintain zero temperature, but
the generalization is straightforward; see, for example,
Ref.~\cite{FURNSTAHL91}.)
The relevant lagrangian density changes to
\beq
   \lagrang (x) \rightarrow {\lagrang}' (x) \equiv
   \lagrang (x) + \mu \Nbar\gamma_0 N \ .
\eeq
The effective action of ${\lagrang}' (x)$ is associated with the
grand potential $\Omega$ of the system, instead of the energy.
The energy follows from
\beq
  E = \Omega + \mu B \ , \qquad
    B \equiv {} - \partial \Omega / \partial \mu \ ,
  \label{eq:EBdefs}
\eeq
where $B$ is the baryon number of the system.

We can again integrate over the baryon fields in the path integral,
just as at zero density.
The result is the fermion determinant at finite density (or chemical
potential), $-i\, \Tr\!\ln K(\mu )$, to which we can add and subtract
the fermion determinant at $\mu = 0$, namely, 
$-i\, \Tr\!\ln K(0)$.
The added term $-i\, \Tr\!\ln K(0)$ cancels the counterterms
exactly as described previously.
Note that it contains the same dynamical scalar and vector fields 
as the fermion determinant at finite $\mu$.
The remaining sum
\beq
 -i\, \Tr\!\ln K(\mu ) + i\, \Tr\!\ln K(0)
 \label{eq:Trlns}
\eeq
is an explicitly density-dependent contribution (it vanishes for 
$\mu = 0$), which is {\em finite}.
(As before, baryon counterterms must be defined when one calculates
beyond one-loop order, as we describe in sec.~\ref{sec:discussion}.)
The scalar potential $U (\phi )$
of Eq.~(\ref{eq:lagrangian}) remains intact; the
only difference is that the scalar (and vector) fields will acquire
different expectation (mean) values due to the presence of valence
(positive-energy) nucleons at finite density.
We emphasize that Eq.~(\ref{eq:Trlns}) applies to \emph{both} the
ground state and RPA excited states.

\section{Ground State}
\label{sec:GS}

We can evaluate the sum Eq.~(\ref{eq:Trlns})
explicitly for the ground state at the one-loop
level to see how the ``no-sea approximation'' arises automatically.
As we stressed earlier, this ``mean-field'' calculation should be
viewed in the context of Kohn--Sham density functional theory.

The mean-field grand potential is defined in the effective-action
formalism by replacing all of
the dynamical meson fields by their mean values, resulting in
\beq
   \int\! dx_0 \, \Omega =  i\, \Tr\!\ln {\Kbar}(\mu ) -
      i\, \Tr\!\ln {\Kbar}(0) + \int\! d^{\, 4} x \, 
      {\Ubar}_m ( {\mathbf x}) \ ,
   \label{eq:omegaone}
\eeq
where the bars indicate that the quantities are to be evaluated
with the static scalar and vector mean fields, which we will
denote by $\phi_0 ({\mathbf x})$
and $\Vzero ({\mathbf x})$.
In particular, the baryon kernel in coordinate space is now
\beq
   \langle x | {\Kbar} (\mu ) | y \rangle
   = \gamma_0 [ \, i \partial_0 + \mu - h ({\mathbf x})]\,
     \delta^{(4)} ( x - y ) \ ,
\eeq
with the single-particle Dirac hamiltonian
\beq
   h ({\mathbf x}) \equiv {}-i \,{\bm \alpha}
   \,{\bm \cdot}\, {\bm \nabla} 
   + \gv \Vzero ({\mathbf x}) 
   + \beta [ M - \gs \phi_0 ({\mathbf x}) ] \ ,
\eeq
where $\beta = \gamma_0$ and 
${\bm \alpha} = \gamma_0 {\bm \gamma}$.

The contribution to $\Omega$
from the mean meson fields is defined by
\beq
   {\Ubar}_m ( {\mathbf x}) \equiv
    \frac{1}{2}\, (\nabla \phi_0 )^2 + U (\phi_0 )
   - \frac{1}{2}\, (\nabla \Vzero )^2
   - \frac{1}{2}\, \mv^2 {\Vzero}^2 
   + \cdots \ ,
    \label{eq:Umdef}
\eeq
where the ellipsis represents
any other polynomial or gradient terms (with finite,
renormalized constants---as yet unknown) that are retained in
the truncated lagrangian to achieve the desired accuracy for 
the nuclear ground state~\cite{FURNSTAHL97,FURNSTAHL00a}.
The equations determining the mean meson fields have not yet
been specified, but they will be determined shortly.

First we observe that $\Kbar (\mu )$ can be diagonalized by
choosing the single-particle basis $\psi_\alpha (x) \equiv
\psi_\alpha ({\mathbf x}) {\mathrm e}^{i \omega x_0}$, where
$\{ \psi_\alpha ({\mathbf x}) \}$ are the normalized 
eigenfunctions of the Dirac equation with eigenvalues 
$E_\alpha$~\cite{HOROWITZ81,SEROT86,SEROT92}:
\beq
  h ({\mathbf x}) \,\psi_\alpha ({\mathbf x}) 
  = E_\alpha \psi_\alpha ({\mathbf x})
  \ , \qquad
  \int\! d^{\, 3} x \, \psi^\dagger_\alpha ({\mathbf x})
  \psi^{\phantom\dagger}_\beta ({\mathbf x}) = 
  \delta_{\alpha\beta} \ .
    \label{eq:KSeigenvalues}
\eeq
This diagonalization works for {\em any} value of $\mu$ and results
in the matrix elements
\beq
  \langle \alpha \omega | \Kbar (\mu ) | \beta \omega' \rangle = 
   2 \pi \,\delta (\omega - \omega' ) \,\delta_{\alpha\beta} \,
   ( - \omega + \mu - E_\alpha ) \ . \label{Eq:matrixel}
\eeq
Applying the appropriate boundary conditions (which reproduce the 
familiar Feynman boundary conditions for free nucleons) 
is equivalent to the
$i \epsilon$ prescription $\omega \rightarrow (1 + i \epsilon )\omega$
for evaluating the baryon kernel~\cite{ITZYKSON80}.
Note, however, that because $h({\mathbf x})$ is to be interpreted
as a Kohn--Sham single-particle hamiltonian, the
eigenvalues $E_\alpha$ have no
directly observable meaning, except at the Fermi surface
$(E_\alpha = \Efermi = \mu)$ \cite{KOHN99}.

Using the result (\ref{Eq:matrixel}) in Eq.~(\ref{eq:omegaone}), we find
\beq
    \Omega (\phi_0 , \Vzero ; \mu) = 
       i \sum_\alpha \int \frac{d \omega}{2 \pi } \ 
        [ \,\ln (-\omega + \mu - E_\alpha)
                 - \ln (-\omega - E_\alpha )]
      + \int\! d^{\, 3} x \, {\Ubar}_m ( {\mathbf x}) \ ,
\eeq
where the sum on $\alpha$ runs over both positive- and
negative-energy eigenvalues.
To evaluate the integrals, one must take care with the analytic
structure of the logarithms, and it is easiest to begin with
the computation of the baryon number, as defined in 
Eq.~(\ref{eq:EBdefs}).
Contour integration produces
\beq
  B = - \partial \Omega / \partial \mu
    = \sum_\alpha \left[ \theta (\mu - E_\alpha ) - 1/2
      \right] \ .
   \label{eq:Bone}
\eeq
To properly define the normal-ordered baryon number, we can use
\beq
     \sum_\alpha \theta (-E_\alpha )
    = \sum_\alpha \theta (E_\alpha )
    = \sum_\alpha \frac{1}{2} \ ,
    \label{eq:itsahalf}
\eeq
which is valid when $E_\alpha = 0$ separates the valence levels
from the Dirac sea.\footnote{%
We assume that at $\mu \not= 0$, as at $\mu = 0$, the positive-energy
(valence) levels are separated in energy from the Dirac sea.
This is true in all practical applications of the QHD lagrangian to
nuclei and nuclear matter.
This is the primary advantage of choosing $\mu = 0$ to 
define the counterterms.}
This result is clearly valid for noninteracting nucleons at
$\mu = 0$, and it remains valid when the valence nucleons are
added and the interactions are turned on, due to the conservation
of baryon number~\cite{SEROT86,FURNSTAHL95,FURNSTAHL97}.

Using Eq.~(\ref{eq:itsahalf}), we can rewrite the
baryon number (\ref{eq:Bone}) as
\beq
     B = \sum_\alpha \left[ \theta (\mu - E_\alpha ) -
      \theta ( -E_\alpha ) \right]
      = \sum_\alpha^{0 \, < \, E_\alpha \, < \, \mu} \!\! 1 \ 
      \equiv \sum_\alpha^{\rm occ} 1 \ ,
   \label{eq:Btwo}
\eeq
where the final sum is over the {\em occupied}
valence orbitals at the given value of $\mu$.
This result is precisely what we should expect from the
boundary-condition prescription discussed earlier, which 
leads to the familiar normal ordering of the baryon-number 
operator~\cite{SEROT86}.

To compute $\Omega$, one must also perform contour integrals,
using care to orient the contours correctly with respect to
the branch points of the logarithms.
The procedure is equivalent to a Wick rotation to
$\nu = -i \omega$ and produces
\beqa
  \Omega (\phi_0 , \Vzero ; \mu ) &=&
      - \sum_\alpha \int \frac{d \nu}{2 \pi } \ 
      [ \, \ln (-i\nu + \mu - E_\alpha )
              -\ln (-i\nu - E_\alpha ) ]
      + \int\! d^{\, 3} x \, {\Ubar}_m ( {\mathbf x}) 
     \nonumber \\
 &=& - \sum_\alpha (\mu - E_\alpha ) \,\theta (\mu - E_\alpha )
      - \sum_\alpha E_\alpha \,\theta (-E_\alpha )
      + \frac{1}{2} \sum_\alpha \mu \ 
      + \int\! d^{\, 3} x \, {\Ubar}_m ( {\mathbf x}) \ ,
     \nonumber \\
 &=& - \sum_\alpha (\mu - E_\alpha )\, [\theta (\mu - E_\alpha )
             - \theta (-E_\alpha )] 
          + \int\! d^{\, 3} x \, {\Ubar}_m ( {\mathbf x}) 
     \nonumber \\[5pt]
 &\equiv& - \sum_\alpha^{\rm occ} (\mu - E_\alpha )
         + \int\! d^{\, 3} x \, {\Ubar}_m ( {\mathbf x}) \ ,
\eeqa
where we have used Eq.~(\ref{eq:itsahalf}) to produce the
normal-ordered result for the baryon number.

By combining the preceding expressions, we can compute the
ground-state energy:
\beqa
  E = \Omega + \mu B &=& 
      - \mu \sum_\alpha [\, \theta (\mu - E_\alpha ) -
                             \theta ( -E_\alpha ) ]
      + \mu \sum_\alpha [\, \theta (\mu - E_\alpha ) -
                            \theta ( -E_\alpha ) ]
     \nonumber \\
& & \quad
    + \sum_\alpha E_\alpha \, \theta (\mu - E_\alpha )
    - \sum_\alpha E_\alpha \, \theta ( -E_\alpha )
      + \int\! d^{\, 3} x \, {\Ubar}_m ( {\mathbf x}) 
     \nonumber \\
&=& \sum_\alpha^{E_\alpha \, < \, \mu} E_\alpha
   - \sum_\alpha^{E_\alpha \, < \, 0} E_\alpha
      + \int\! d^{\, 3} x \, {\Ubar}_m ( {\mathbf x})
    \label{eq:thenub} \\[5pt]
&=&  \sum_\alpha^{\rm occ} E_\alpha
        + \int\! d^{\, 3} x \, {\Ubar}_m ( {\mathbf x}) \ .
    \label{eq:noseaE}
\eeqa
We emphasize that the final sum over occupied valence
states only is not the result of a {\em vacuum} subtraction,
since the trace with $\mu = 0$ [which produces the second
sum in Eq.~(\ref{eq:thenub})] still contains the 
self-consistent mean fields $\phi_0( {\mathbf x})$ and $\Vzero( {\mathbf x})$.
The true vacuum subtraction was performed earlier when we
derived the renormalized (and finite) ${\Ubar}_m$ in
Eq.~(\ref{eq:Umdef}).

Thus we have arrived at the ``no-sea approximation'' for the ground
state.
The energy is determined by a sum over valence-orbital eigenvalues
and by a local potential in the meson fields (and their derivatives)
with finite, but unknown, constants.
How can we see that the vacuum dynamics is still included?

\begin{figure}[t]
\begin{center}
\includegraphics*[width=5.2in,angle=0]{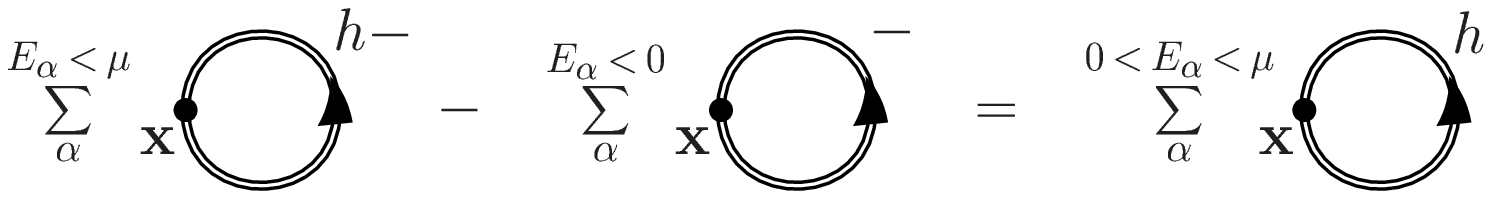}

\caption{Diagrammatic representation of the subtraction 
procedure described in the text, which yields the ``no-sea
approximation'' for the ground-state scalar density 
$\protect\rhos ({\mathbf x})$.
(Local meson terms are not shown.)}
\label{fig:rhos}
\end{center}
\end{figure}

Let us recall how we arrived at Eq.~(\ref{eq:noseaE}).
We first showed that at $\mu = 0$, the fermion determinant could
be written as a generalized derivative expansion in 
local terms [Eq.~(\ref{eq:Kzero})] and could therefore be
canceled exactly by the counterterms present in the 
original lagrangian.
The physics behind this is that in an EFT dealing with long-range
dynamics, the vacuum contributions are
so poorly resolved that they can be
accurately represented by local terms in a derivative
expansion containing the meson fields~\cite{FURNSTAHL90}.
At finite $\mu$, we then added and subtracted the
$\mu = 0$ determinant from
the grand potential; the added term cancels exactly against the
counterterms (by construction), while the subtracted term was 
{\em rewritten} as a sum over (negative) eigenvalues to reveal 
that it exactly removes
these contributions from the first term in Eq.~(\ref{eq:thenub}), 
which originates from $\Tr\!\ln {\Kbar} (\mu )$.
In principle, we could have argued that the negative-energy
part of the first sum in Eq.~(\ref{eq:thenub})
could indeed be represented by local terms in the fields and derivatives
(since such vacuum contributions are not resolved in the low-energy EFT)
and simply canceled them away exactly by the local counterterms.
Instead, we inserted the intermediate step involving 
$\Tr\!\ln {\Kbar}(0)$ to show explicitly
the local nature of the second sum in Eq.~(\ref{eq:thenub})
and that it is equivalent to local counterterms.
The analogous procedure that produces the nucleon scalar
density is shown schematically in Fig.~\ref{fig:rhos}.

All that remains in the energy $E$ is the local meson potential
and a sum over valence-orbital eigenvalues; the former
has the same form as the counterterms, but with finite, 
unknown constants that will ultimately get fitted to data
to encode the true QCD vacuum dynamics into the energy.
As long as we fit the parameters to empirical data, none of the vacuum
or short-range QCD dynamics is omitted~\cite{FURNSTAHL97,SEROT97}.

The reason these subtraction procedures work is that 
the local terms in the derivative expansion
[e.g., Eq.~(\ref{eq:Kzero})] have the same form for 
{\em any} values of the fields $\phi$ and $V^\mu$.
The counterterm parameters (constants)
multiplying the fields are fixed for a given level
of approximation to the original lagrangian.
Thus at $\mu = 0$, the counterterm subtraction can be implemented
while the meson fields are still dynamical (i.e., integration
variables in the path integral), while at $\mu \not= 0$, a similar
set of counterterm subtractions produces finite results for
the grand potential or energy of the system as a function of
the mean meson fields.
The nuclear energy at the one-loop level automatically reproduces 
the well-known ``no-sea approximation'', as discussed above.

This discussion also reveals why the choice of $\mu = 0$ for the
original normalization of the QHD lagrangian is so convenient.
Since the positive- and negative-energy eigenvalues are
separated by $E^0_\alpha = 0$ at $\mu = 0$ and remain separated
by $E_\alpha = 0$ at finite $\mu$ (at least in all nuclear structure
applications that we are aware of), the final expression for the
energy [Eq.~(\ref{eq:noseaE})]
contains a sum over valence nucleon orbitals only.
Thus the ``no-sea approximation'', which still allows the inclusion
of vacuum dynamics through the fitted parameters in the local meson
potential, arises naturally in QHD because of a 
{\em convenient} choice for the normalization of the lagrangian.
Other choices ($\mu \not= 0$) for this normalization are certainly
possible, but if one chooses a value of $\mu$ for the initial
vacuum subtraction that will ultimately lie {\em within} the
spectrum of positive-energy eigenvalues, subsequent subtractions
to define the ground-state energy will be more complicated.
Such a renormalization procedure could be implemented in principle,
but it would be messy, and the extra complication is unnecessary;
the most convenient choice is $\mu = 0$, which leads naturally
to familiar results for the one-loop ground-state energy.

To determine the mean meson fields, one extremizes the expression
for the energy [Eq.~(\ref{eq:noseaE})] with respect to these fields.
This leads (in general) to nonlinear differential equations for
the meson fields, with the nucleon scalar and baryon densities
as the sources.
(See, for example, Refs.~\cite{SEROT92,FURNSTAHL95,FURNSTAHL97}.)
Thus the mean meson fields and the nucleon wave functions must
be determined {\em self-consistently}, as is well 
known~\cite{MILLER74,BROCKMANN77,HOROWITZ81}.
We emphasize, however, that the mean meson fields and the nucleon
densities are all {\em local}, time-independent
functions of a single spatial variable $\mathbf x$.
Moreover, since the meson parameters are fitted to empirical
many-body data, the meson fields are to be interpreted as
relativistic Kohn--Sham potentials.

\section{Linear Response (RPA)}
\label{sec:RPA}

The 
renormalization procedure detailed in sec.~\ref{sec:EFT}
also defines the linear response
of the ground state to time-dependent fields (RPA).
To see the consequences of this procedure, 
we consider the effective action with fluctuations
around the static, ground-state fields:
\beq
   \phi(x) = \phi_0({\bf x}) +  \wt\phi(x) \ , 
    \qquad 
    V_\mu(x) = V_0({\bf x})\,\delta_{\mu 0} + \wt V_\mu(x) \ .
     \label{eq:fluctuations}
\eeq
The fluctuations are denoted with tildes and are explicitly
of $O( \hbar^{1/2})$.
We will work to leading order in the fluctuations, which turn
out to be $O(\hbar )$ and yield the familiar 
RPA~\cite{DAWSON90,FETTER71,CHIN77,NEGELE88,WEHRBERGER90}.

Note that the ground-state mean fields are determined by
extremizing the energy [Eq.~(\ref{eq:noseaE})], which implies
that we are working in the canonical ensemble at fixed baryon
number (or density).
The desired density can be imposed by applying the appropriate
infinitesimal boundary conditions on the baryon 
propagator~\cite{SEROT86,DAWSON90,PIEKAREWICZ01}.
A careful analysis based on the grand (thermodynamic) potential
produces identical results, when one works at the level of the
lowest-order RPA, as we are here~\cite{KOHN60,WEHRBERGER90}.
In particular, the RPA propagators are to be evaluated in the
presence of the ground-state mean fields $\phi_0$ and $V_0$.

Since the effective action $\Gamma$ is
the generator of one-particle-irreducible (1PI) Green's
functions, all we need to describe the linear response are the 
terms in $\Gamma$ that are quadratic in the fluctuation fields.
These terms give us the meson polarization insertions, expressed
in terms of baryons propagating in the static mean fields.
(Recall that terms {\em linear} in the fluctuation fields vanish
identically, since their coefficients are zero by virtue of the
ground-state field equations.)
In particular, $\partial^2\Gamma/\partial\wt\phi(x)\partial\wt\phi(y)$
evaluated at $\phi_0({\bf x})$ and $V_0({\bf x})$ is the inverse
scalar meson propagator in the presence of the mean fields.
The inverse vector meson propagator and other response functions
follow similarly. 

It is critical to recognize that the fermion determinants 
$\,i\Tr\!\ln K(\mu )$ and all
of the local meson terms retain {\em exactly the same form} as in
the ground-state calculation.
In particular, the numerical constants in these counterterm contributions are
determined from fits to bulk nuclear properties, 
and the {\em same} constants
are used in the RPA calculation; all that changes are the values
of the meson fields due to the fluctuations.

Thus we can make the same substitutions and subtractions as we
made in considering the ground state.
We therefore carry out the expansion of $\Gamma$, focusing
on the $i\Tr\!\ln K(\mu)$ term and substituting $-i\Tr\!\ln K(0)$
for the local counterterms.  
[Note that there are no ``bars'' on these $K$'s, since they
contain the modified fields of Eq.~(\ref{eq:fluctuations}).]
If we consider the $\wt\phi$ terms in a schematic notation,
we find that the quadratic contribution
\beq
  \ln(\GKS^{-1} + \gs\wt\phi)
           = \ln(\GKS^{-1}) \bigl[\,\ldots  
      {}- \frac{1}{2} \,\GKS\, \gs\wt\phi\, \GKS\,
              \gs\wt\phi
                     + {} \ldots\,\bigr]
\eeq
gives the lowest order RPA ring diagram 
$[\tr\,\GKS(x,y)\,\GKS(y,x)]$ for the scalar meson polarization
insertion,
where ``tr'' denotes a trace over spin and isospin.
Here the ring is formed from the ground-state
baryon propagators $\GKS$, which involve the
mean meson fields and the resulting self-consistent Kohn--Sham
(KS) baryon wave functions from Eq.~(\ref{eq:KSeigenvalues}), 
as discussed earlier.
The final RPA equations will be the same as in 
Refs.~\cite{PIEKAREWICZ00,PIEKAREWICZ01},
so we focus here on the spectral content of the nucleon part rather than
the details of the derivation.

\begin{figure}[t]
\begin{center}
\includegraphics*[width=4.2in,angle=0]{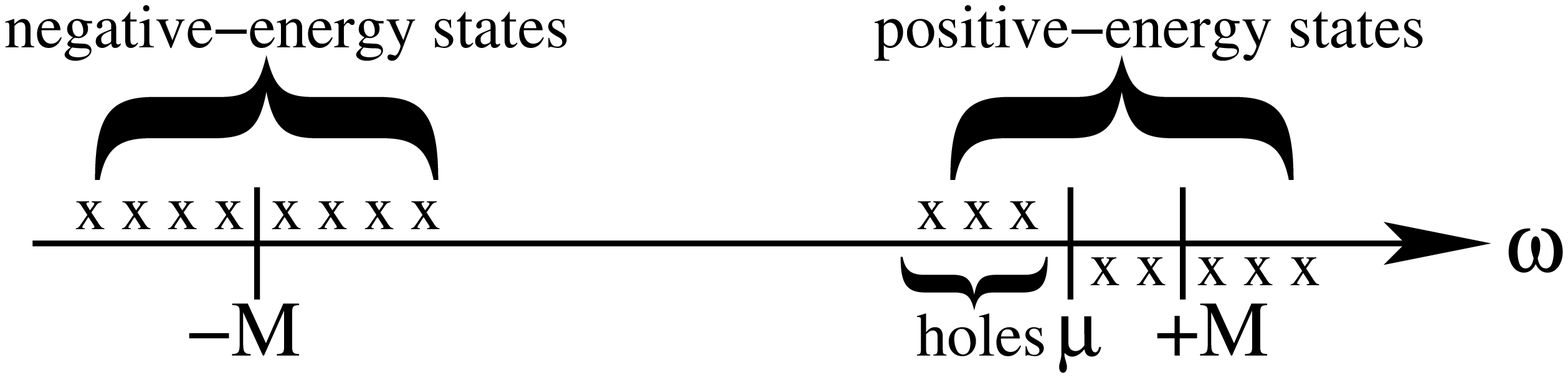}

\caption{Spectral content of the Kohn--Sham propagator,
    which is the same as the nucleon propagator in a 
     relativistic Hartree approximation~\protect\cite{SEROT86}. 
     Single-particle
     states with energies between $-M$ and $+M$ are bound,
     while those with energies less than the chemical 
     potential $\mu$ are occupied.}
\label{fig:poles}
\end{center}
\end{figure}

Since the baryon propagator can be written as a bilinear form
using the Kohn--Sham wave functions (and their adjoints), we
can use the eigenvalue equation (\ref{eq:KSeigenvalues}) to
convert the propagators to frequency space.
They take the form
\beq
  \GH({\bf x},{\bf x'};\omega) = \sum_{\alpha}
  \left[ 
  {\cal U}_\alpha({\bf x})\,\overline{\cal U}_\alpha({\bf x'})
  \left(
   \frac{\theta(E^+_\alpha-\mu)}
        {\omega-E^+_\alpha+i\eta}
    +
   \frac{\theta(\mu-E^+_\alpha)}
        {\omega-E^+_\alpha-i\eta}
  \right) +
  \frac{{\cal V}_\alpha({\bf x})\overline{\cal V}_\alpha({\bf x'})}
        {\omega-E^-_\alpha-i\eta} 
  \right] \;,
 \label{eq:GH}
\eeq
where ${\cal U}_\alpha({\bf x})$ and ${\cal V}_\alpha({\bf x})$
are positive- and negative-energy energy solutions, respectively,
of a Dirac equation with background scalar $\phi_0({\bf x})$ 
and vector $V_0({\bf x})$ fields:
\beq 
  \{ -i\,{\bm \alpha}
   \,{\bm \cdot}\, {\bm \nabla} 
      + \beta [M - \gs\phi_0({\bf x}) ] 
      + \gv V_0({\bf x}) \}
  \left\{
     \matrix{{\cal U}_{\alpha}({\bf x}) \vphantom{\Big)} \cr
             {\cal V}_{\alpha}({\bf x}) \cr}
  \right\} 
      =  
  \left\{
     \matrix{E^+_\alpha\,{\cal U}_{\alpha}({\bf x}) 
                     \vphantom{\Big)} \cr
             E^-_\alpha\,{\cal V}_{\alpha}({\bf x}) \cr}
  \right\} 
  \label{eq:DiracEq} \;,
\eeq  
and $\eta$ is a positive infinitesimal.
Note that including additional background fields does not affect 
the present discussion.
The pole structure of Eq.~(\ref{eq:GH}) is illustrated in
Fig.~\ref{fig:poles}.
(The infinitesimals that enforce the appropriate boundary
conditions are defined at both zero and finite density in the
usual way~\cite{DAWSON90}).

The frequency integration over the two $\GKS$ propagators 
picks up contributions from 
both particle-hole ($ph$) pairs and 
particle-negative-energy ($p/-$) pairs for the determinant
with $\mu >0$, where ``particle'' implies a Dirac 
state that is unoccupied in the ground state, while
``hole'' implies an occupied state.
In contrast, for $\mu = 0$, we get contributions from
{\em all} positive-negative-energy ($+/-$) terms,
which resemble the usual ``vacuum polarization''.
Remember, however, that all of these terms involve the 
Kohn--Sham background fields, since the true vacuum 
subtraction was made back in sec.~\ref{sec:EFT}.

\begin{figure}[t]
\begin{center}
\includegraphics*[width=5.2in,angle=0]{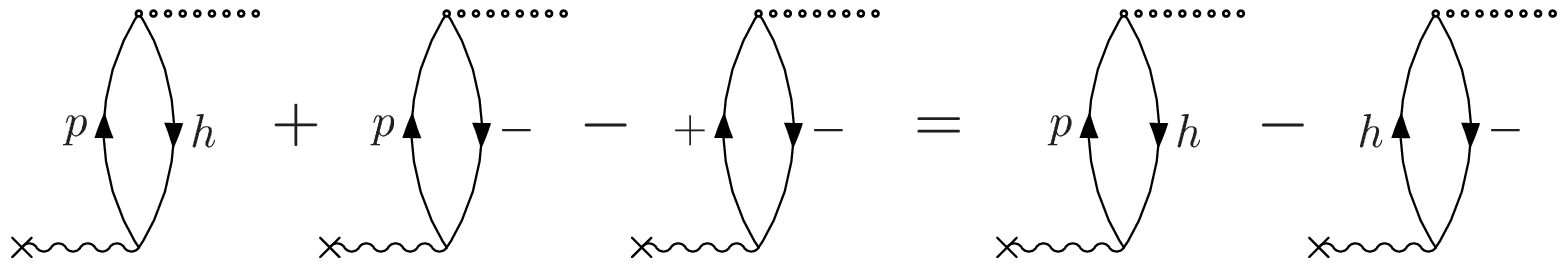}

\caption{The subtraction procedure for RPA in the ``no-sea''
prescription (represented diagrammatically) 
yields particle-hole ($ph$) pairs minus mixing terms between
occupied positive-energy and negative-energy basis 
states $(h/-)$.
The negative-energy states are important for maintaining the
completeness of the Dirac single-particle basis.}
\label{fig:rpaone}
\end{center}
\end{figure}

When all of these ring contributions are combined, the net result: 
$(ph)+(p/-)-(+/-)$ is just the difference $(ph) - (h/-)$,
as illustrated in Fig.~\ref{fig:rpaone}.
Thus, by carrying out our consistent normalization
and renormalization
procedures in the QHD EFT, the RPA response contains
{\em both} familiar particle-hole pairs {\em and}
mixing between occupied particle states and
negative-energy states in the single-particle (Kohn--Sham)
basis.
Because this second term forces the inclusion of 
negative-energy states that complete the Dirac basis,
it is crucial for maintaining Thouless' theorem and 
various conservation 
laws~\cite{DAWSON90,FURNSTAHL87,PIEKAREWICZ00,PIEKAREWICZ01}.
We have therefore succeeded in showing that the standard
rules of quantum field theory (and a convenient choice for
the normalization of the QHD lagrangian) lead to the RPA
contributions in Fig.~\ref{fig:rpaone}.

Why should we expect that this RPA framework yields reasonable
results for nuclear collective excitations?
The first thing to remember is that the Kohn--Sham energy
functional, while an {\em approximation} to the exact energy
functional, is an excellent approximation over the density
regime of interest and is fit not only at the equilibrium
point but also in the vicinity of this point.
This (approximate) energy functional goes beyond
simple Hartree theory and implicitly includes contributions from
nucleon exchange, correlations, hadronic structure, 
short-distance physics, and the quantum vacuum into the
description of nuclear densities and energies~\cite{SEROT01}.
This description is very accurate, based on the agreement
between calculated ground-state results and the 
data~\cite{FURNSTAHL95,RING96,FURNSTAHL97}.
Since the ground-state energy functional is fitted
to the empirical bulk properties of nuclei, it includes all
of the relevant long-range physics.
Just as we would expect a modest extrapolation in density or
proton fraction away from the nuclear matter
equilibrium point to be accurately described by the fitted 
energy functional, we expect that low-lying
(acoustic) collective excitations of the mean-field ground
state should be described accurately as well.
Nevertheless, one must include {\em all\/} of the states in 
the complete Dirac basis to achieve this accuracy.

Although recent RPA 
calculations~\cite{PIEKAREWICZ00,PIEKAREWICZ01,RING00,RING01,RING01a}
include only a subset of the local meson terms (cubic and quartic
non-derivative scalar terms) present in the full effective 
potential, power 
counting~\cite{FURNSTAHL97,SEROT97} implies that other terms,
such as mixed scalar--vector cubic and vector quartic terms,
are equally important.
Calculations with these additional terms have yet to be done,
and data on excited nuclear states may provide new constraints
that can help determine the numerical coefficients of these
terms in the energy functional.

\section{Relationship Between EFT and Earlier Approaches}
\label{sec:Dirac}

As we have seen in the preceding EFT discussion,
nucleons in the occupied Fermi sea modify the QCD 
vacuum, which in turn acts back on the valence nucleons through
the fitted terms in the local meson potential~\cite{FURNSTAHL97}.
If one constructs a \emph{renormalizable} hadronic field theory based
on low-energy degrees of freedom, one is making an explicit
model for this vacuum dynamics; namely, that loop integrals
involving these hadronic degrees of freedom can adequately describe
the vacuum~\cite{WALECKA74,SEROT86}.
For example, in the relativistic Hartree 
approximation (RHA), the contribution of the Dirac sea
for a uniform system is a Casimir energy at finite density.
This arises from the nonzero scalar mean field $\phi_0$, 
which changes the sum over energies in the Dirac sea:
\beq    
  \delta H = {} - \sum_{{\bf k}\lambda}\, \bigl[
    ({\bf k}^2 + \Mstar{}^2)^{1/2} - ({\bf k}^2 + M^2)^{1/2}
    \bigr]  \ ,
\eeq
where $\Mstar \equiv M - \gs \phi_0$, and $\lambda$ is the spin-isospin
degeneracy.
After renormalization, the resulting contribution to the 
ground-state energy is an (infinite) polynomial
in $\phi_0$~\cite{SEROT86}. 
The phenomenological consequences (after conventional 
renormalization) include smaller 
mean fields than those found in best-fit mean-field theories; the
smaller fields lead to incorrect spin-orbit splittings in 
nuclei~\cite{SEROT86,SEROT92,FURNSTAHL97,FURNSTAHL98}.

In phenomenologically successful covariant mean-field 
models~\cite{RUFA88,REINHARD89,GAMBHIR90,RING96}, these
Dirac sea contributions are neglected by fiat.
This means that the nucleon scalar density, for example, 
is computed from a sum over self-consistent, positive-energy, 
single-particle states only. 
Yet the self-consistent Hartree propagator 
{\em must\/} be 
constructed from a {\em complete} basis of Dirac single-particle 
wave functions; otherwise, it fails to satisfy the 
appropriate differential equation~\cite{PIEKAREWICZ00}.
Thus
the single-particle Hartree propagator takes the 
form of Eq.~(\ref{eq:GH})~\cite{FURNSTAHL85,NISHIZAKI86}, in which 
the sum over states in Eq.~(\ref{eq:GH}) includes 
negative-energy solutions to the Dirac equation.

For ground-state quantities, however, the ``no-sea approximation''
implies that one should drop the contributions from the 
negative-energy poles  of this Green's function. 
Although the ``no-sea approximation'' is known 
to be both covariant and thermodynamically 
consistent~\cite{SEROT86}, its early use
lacked any formal justification, 
and it was not clear how to generalize 
it~\cite{REINHARD89,GAMBHIR90,RING96}. 
Moreover, its accuracy was suspect, especially in view of the
unnaturally large contributions from the Casimir
energy~\cite{FURNSTAHL97b}.
Our discussion in the preceding sections (based on modern EFT)
shows how this approximation arises naturally from the standard
rules of quantum field theory, {\em without an explicit model
for the dynamics of the vacuum} and without dropping any poles.
Extension to higher-order approximations is straightforward, 
but tedious~\cite{HU00}, as we will comment on in the 
Discussion section below.

When extended to linear-response theory, the ``no-sea
approximation'' leads to the naive expectation
that only particle-hole pairs should be used in the 
RPA configuration space~\cite{MA97,MA97a,MA99}. 
It has been known for fifteen 
years~\cite{FURNSTAHL87,DAWSON90},
however, that the consistent linear response to 
a ``no-sea'' ground state must include, in addition to the 
conventional particle-hole ($ph$) pairs, contributions that mix
occupied states or holes ($h$) and negative-energy ($-$) 
basis states from the Hartree ground-state calculation. 
This requirement has been 
discussed using perturbation theory and Green's functions 
for the elastic case \cite{FURNSTAHL87}, a self-consistent,
conserving-approximation functional approach~\cite{DAWSON90}, 
and a time-dependent Hartree approximation~\cite{RING01a}. 
Neglecting these contributions has disastrous 
phenomenological consequences. 
Chief among them is the failure to fully remove 
the spurious center-of-mass strength from the isoscalar dipole 
response as well as the violation of electromagnetic current
conservation. 
 
The inclusion of ($h/-$) contributions in an RPA calculation can
be accomplished explicitly by expanding the RPA matrix to include 
these additional configurations, in what is called the spectral 
RPA approach~\cite{DAWSON90}, or by using a more efficient non-spectral
approach (and looking for poles in the response 
function)~\cite{PIEKAREWICZ00,PIEKAREWICZ01}.
A formal solution to the consistent use of Eq.~(\ref{eq:GH}),
proposed in Ref.~\cite{DAWSON90}, is to shift (by fiat) the
negative-energy poles to the lower-half plane (see also
Ref.~\cite{RING01a}).  Then one picks up the
desired poles for both ground states and excited states.
This shift is equivalent to normal ordering the single-particle
density operator.

The explanations for including ($h/-$) contributions are not at all
satisfying when based on earlier approaches that either modeled the
QCD vacuum as an interacting Dirac sea of nucleons or that neglected
the Dirac sea completely in the computation of the Hartree ground
state.
The primary motivation for including these ($h/-$) terms was that
they were necessary to maintain the conservation laws and to
agree with the observed phenomenology.
Within the modern EFT/DFT/KS approach, however, one realizes that 
the long-range (nucleon and meson) degrees of freedom {\em cannot}
adequately describe the QCD vacuum.
Nevertheless, since the vacuum contributions from these terms can
always be represented in the EFT by local terms in a meson potential, 
they can be combined with other terms in the QHD lagrangian 
(which contains all nonredundant terms consistent with the 
symmetries), leading to a meson effective potential that contains 
all vacuum effects when it is fitted to empirical data.

Moreover, a straightforward extension of the ground-state calculation
to excited states, which is performed by allowing the meson fields
to fluctuate around their mean values, naturally leads (again from
the standard procedures of quantum field theory) to all the 
necessary loop contributions in the RPA calculation.
This derivation shows that the vacuum has {\em not} been neglected;
the parts of the vacuum that are beyond the limits of description by
the EFT are parametrized in terms of the meson potential, and
the long-range vacuum modifications arising from the $(h/-)$ mixing
are included explicitly, as they should be.
Thus the modern approach not only vindicates the inclusion of the
correct terms in the RPA linear response, but it also shows why
certain terms (the so-called $N \Nbar$ pairs) should not be
calculated explicitly, because they are beyond the range of validity
of the EFT, and they will always be included {\em implicitly} in
the meson potential present in the mean-field hamiltonian.

\section{Discussion and Summary}
\label{sec:discussion}

The purpose of this paper is to show how the modern EFT/DFT
approach to QHD~\cite{SEROT97,FURNSTAHL00} can be 
implemented straightforwardly at the mean-field level 
using standard quantum-field-theory procedures.
The results imply that the so-called ``no-sea approximation''
for the nuclear ground state and its generalization to
the RPA linear response (which includes long-range
contributions from
negative-energy states in the complete Dirac single-particle
basis) are justified by modern field-theoretic
approaches to the nuclear many-body problem.
Since one expects that QCD vacuum physics cannot be adequately
described by low-energy hadronic degrees of freedom alone, 
vacuum contributions from these terms must be subtracted away
by the counterterms present in the QHD lagrangian and combined
with local meson terms that are explicitly fitted to empirical
data.
As we emphasized, only the {\em sum} of all of these terms is
constrained by experimental results, and by fitting to data we
can implicitly include the vacuum effects, as well as other
short-range and many-body effects~\cite{FURNSTAHL97}.

The implicit ``no-sea'' subtraction procedure removes contributions
from explicit sums over the entire Dirac sea of nucleons, which 
are beyond the realm of the low-energy EFT anyway.
This means that no explicit calculations of loop 
integrals or counterterms must be made (unlike the
Relativistic Hartree Approximation \cite{SEROT86,SEROT92}).  
In principle, an infinite set of counterterms is needed to describe
the vacuum dynamics, but in practice,
the finite residual parts are under-determined, and naturalness implies
that most are numerically unimportant~\cite{FURNSTAHL00a}.
Moreover, the mean-field computation of the single-particle Dirac
wave functions can be related to relativistic Kohn--Sham theory
using density-functional arguments, which show that more than
simple single-particle (``Hartree'') physics is included in these
wave functions.
The key to these arguments is that the local meson potentials provide
an {\em excellent approximation} to the exact ground-state energy 
functional in the density regime of 
interest~\cite{FURNSTAHL97,SEROT01}.
Thus mean-field predictions for bulk nuclear properties will
be accurate, and the theory is predictive when the most important
(dominant) local terms have been fitted to data.

In summary, we regard these as our most important conclusions:
\vspace{-0.10in}
\begin{enumerate}
\item{}
 The strength of the EFT is that while the
 short-distance behavior of the theory is (probably) incorrect, 
 it can be corrected systematically by the counterterms.  
 The well-known ``no-sea approximation'' is a particular, yet convenient,
 prescription for performing this renormalization. 
 It is therefore
 incorrect to view the ``no-sea approximation''
 as an ``empty'' Dirac sea.  
 Indeed, not only have vacuum effects been included, but
 the true vacuum dynamics becomes encoded in a small number of
 empirical constants that define the local meson potential.
\item{}
 The successful phenomenology of the ``no-sea approximation''
 is justified by density-functional theory. 
 The self-consistent Kohn--Sham
 equations contain Hartree theory as a particular limit, 
 yet they go beyond Hartree theory by implicitly including 
 many-body correlations, nucleon exchange,
 and short-range effects. 
 This is achieved by fitting the empirical constants in the 
 model to bulk 
 properties of nuclei, rather than to two-body data.
\item{}
 The consistent linear response of the ``no-sea'' 
 ground state must
 include contributions that mix (positive-energy) holes and and 
 negative-energy states,
 in addition to the familiar particle-hole excitations. 
 This result is
 dictated by the EFT that demands the same renormalization 
 scheme for
 the ground state as for the linear response (excited states).
 Moreover, when the Kohn--Sham 
 fields determined for the ground state
 are also used for the computation of the excited states, 
 fundamental conservation laws are maintained, 
 thereby guaranteeing the
 phenomenological success of the RPA. 
 Equivalently, by demanding that
 the particle-hole interaction driving the RPA response be 
 consistent with the accurately calibrated particle-particle 
 interaction used to
 generate the ground state, the small fluctuations about 
 equilibrium are guaranteed to be accurately reproduced.
 This is true even though existing RPA
 calculations within the EFT/DFT approach include only
 scalar cubic and quartic terms, which are just a subset of all 
 allowed terms, but which
 are sufficient for quite a good description of ground-state 
 nuclear properties. 
\end{enumerate}

Finally, although our discussion has thus far
been entirely in the context
of the one-loop approximation to the QHD effective action (or density
functional), 
the underlying principles are more general.
One can improve the analysis to include long-range 
correlations more explicitly within an EFT/DFT framework,
by exploiting the effective-action formalism.
Explicit, dynamical, long-range terms are expected to introduce
new nonanalytic functions of the nucleon densities,
which should improve the approximation to the exact energy
functional and allow for extrapolation outside the density
regime described accurately by the mean-field treatment 
discussed above.

Calculations beyond one-loop order have been studied by
Hu~\cite{HU00}.
Several important new features arise: first, one must retain 
the baryon counterterms in the lagrangian, since many of the
vacuum contributions involve expansions in local terms with
these forms.
Second, one must use care in removing redundant terms from the
lagrangian~\cite{FURNSTAHL97,SEROT97} and develop a systematic
procedure for redefining the fields, so that order-by-order
in the relevant expansion parameter ($\hbar$, or ``hole-lines'',
etc.), the lagrangian can always be recast in a standard
(``canonical'') form.
This allows vacuum subtractions to be made unambiguously as
one proceeds to more and more complex approximations.
Third, as shown by Hu for a wide class of approximations beyond
one-loop order, it is always possible to separate the long-range
nucleon loops from the short-range and vacuum parts of the
loop integrals.
The latter can be written in the same form as local
counterterms and treated analogously to the subtracted terms
described above; thus, one never has to calculate explicitly
either the short-range terms or the counterterms that cancel them.
The long-range terms must be retained and calculated explicitly;
these resemble familiar nuclear many-body integrals, like nucleon
exchange, rings, ladders, etc.~\cite{FETTER71}.

Thus, in the end, we have a systematic way to generalize the
relativistic nuclear many-body problem beyond one-loop 
order~\cite{SEROT97,HU00}.
Long-range nuclear terms can be organized and calculated in much
the same way as in conventional nuclear structure physics; the
only differences are that we now have four-component Dirac wave
functions and propagators, the meson propagators are retarded
and mixed together by various terms in the meson potential, and
there are a small number of unknown constants that specify the
short-range and vacuum QCD behavior, which must be determined by
fitting to many-body (or nuclear matter) data.
These more sophisticated analyses of effective QHD 
lagrangians will provide the basis for future investigations.

\vspace*{-2pt}

\acknowledgments

We thank H.-W.\ Hammer, R.~J.\ Perry, and J.~D.\ Walecka for useful comments.
This work was supported in part by the National Science Foundation
under Grant Nos.~PHY--9800964 and PHY--0098645, and by the 
Department of Energy
under Contract Nos.\ DE--FG02--87ER40365, and DE--FG05--92ER40750.

\end{document}